\documentclass[a4paper,fleqn,usenatbib]{mnras}
\usepackage{amssymb,amsmath,graphicx,psfrag,mathtools}
\usepackage[T1]{fontenc} 
\usepackage{ae,aecompl} 
\usepackage{url}
\usepackage{revsymb}
\hypersetup{draft} 
\title[Symmetry Breaking in Repeating Fast Radio Bursts]{Symmetry Breaking
in Repeating Fast Radio Bursts}
\author[J. I. Katz]{
J. I. Katz,$^{1}$\thanks{E-mail katz@wuphys.wustl.edu} 
\\
$^{1}$Department of Physics and McDonnell Center for the Space Sciences,
Washington University, St. Louis, Mo. 63130 USA 
}
\date{Accepted XXX.  Received YYY; in original form ZZZ} 
\pubyear{2022} 
\date{\today}
\begin{document} 
\label{firstpage} 
\pagerange{\pageref{firstpage}--\pageref{lastpage}} 
\maketitle 
\begin{abstract}
	Repeating Fast Radio Bursts show temporal symmetry breaking on
	millisecond time scales (the ``sad trombone'').  On a time scale of
	days the repetitions of FRB 180916B occur at frequency-dependent
	phases of its 16.3 d period.  Some models predict that all such
	periodic repeating FRB have the same sign of temporal asymmetry,
	while others predict that sources with both signs are equally
	abundant.  Future observations of other periodically modulated
	repeating FRB may distinguish among models on this basis.
\end{abstract}
\begin{keywords} 
radio continuum, transients: fast radio bursts, accretion, accretion discs,
stars: binaries: close
\end{keywords} 
\section{Introduction}
Observations of repeating Fast Radio Bursts (FRB) indicate temporal symmetry
breaking on very different time scales:
\begin{itemize}
	\item On times ${\cal O}(\text{1 ms})$ repeating FRB frequently
		show a ``sad trombone'' effect: during a burst, emission
		slides to lower frequencies.  This is distinct from the
		source's plasma dispersion \citep{ZP21,B22}.  No systematic
		``happy trombone'' slide to higher frequencies has been
		observed in any FRB.
	\item On times ${\cal O}(\text{1 d})$ the periodically modulated
		activity of FRB 180916B (the only firmly confirmed periodic
		FRB, and the only one observed in detail) occurs at earlier
		phases of its 16.3 d period at higher frequencies than at
		lower frequencies.  This was discovered at lower frequencies
		\citep{PM21,ZP21} and \citet{B22} extended it to a third
		frequency band, suggesting a monotonic trend.
\end{itemize}

The extensive sky coverage of CHIME/FRB \citep{C21} is rapidly increasing
the number of known repeating FRB.  The activity of some of these sources
may be periodically modulated like that of FRB 180916B (there is no evident
reason why this object should be exceptional), although insufficient data
have yet been gathered to establish this.  It will be determined whether all
such sources have the same dependence of active phase on frequency, whether
some sources show one sign of frequency dependence and others the other sign
(and with what comparative abundance), or whether the activity phase of some
periodic FRB is independent of the radio frequency of the observations.

The analogous question can be asked of ``trombone'' effects.  There may
already be sufficient data to establish that ``sad trombones'' are
ubiquitous in repeating FRB, and that ``happy trombones'' may not occur at
all.
\section{Symmetry and Symmetry Breaking}
\label{symmetry}
Temporal symmetry breaking may distinguish among models of FRB behavior.
For example, plasma refraction is greater at lower frequencies, so that
lower frequency radiation follows a propagation path between plasma lenses
that deviates from the straight line path of unrefracted radiation more than
does higher frequency radiation.  If there are plasma lenses on the
propagation path then lower frequency radiation arrives later because of its
greater path length, even if the lenses are thin and almost all the path is
effectively in vacuum.  If this effect is the cause of the frequency
dependence of phases of periodically modulated FRB then their phases will be
earlier at higher frequencies in every periodically modulated FRB source.

In contrast, if FRB, either individual pulses or the active window of
periodic activity, result from the sweep of a beam or a beam envelope
across the observer's direction, then in the FRB population both senses of
time dependence occur equally often because there is no preferred direction
of sweep of a beam on the sky (or of the cross-product of its rotation
vector with the direction to the observer).  This applies to almost any
observable, including polarization angle, variation of rotation measure,
sawtooth intensity variation, {\it etc.\/}

Over the FRB population one sense of rapid ($\sim 1\,$ms) variation (the
``sad trombone'') is preferred, excluding sweeping of a steady narrow beam
across the observer's direction as the origin of the bursts.  No such
statement can be made at present about the chromaticity of phase in
periodically modulated FRB activity because it has been observed in only one
source \citep{B22}.

An individual repeating FRB may show the same sense of temporal asymmetry in
pulse after pulse, or (if periodically active) in active period after active
period; a sawtooth pattern may sweep across the observer's direction time
and time again.  In some models there is no relation between our direction
and the direction of sweep through a sawtooth, so that some repeating FRB
sources show systematically positive time-skewness $F(\tau)$, closely
related to the bispectrum, of an observable $f(t)$ \citep{HMM63,FK67,WSKC78}
\begin{equation}
	\begin{split}
		F(\tau) \equiv &\left\langle
		\left(f(t) - \langle f \rangle_t \right)^2
		\left(f(t+\tau) - \langle f \rangle_t \right)\right\rangle_t\\
		&- \left\langle \left(f(t) - \langle f \rangle_t \right)
		\left(f(t+\tau) - \langle f \rangle_t \right)^2
	\right\rangle_t
	\end{split}
\end{equation}
while other sources show $F(\tau)$ systematically negative.

Rotating beams with a sawtooth angular dependence are equally likely to have
positive as negative time skewness if there is no preferred sense of
rotation with respect to the beam pattern.  This is plausible for slow
rotation or precession of a compact object, but not necessarily for fast
rotation that may influence the emission mechanism or if the rotation of the
beam is produced by orbital motion that also determines the radiated power
and spectrum.

In other models there is an intrinsic temporal asymmetry that all observers
see; decaying eruptions are examples, consistent with the sad trombone
effect.  Once a sufficient number of repeating FRB with periodically
modulated activity are observed, this test may distinguish among models.
This is distinct from tests \citep{K21} based on the rate of change of
period of a single periodically modulated repeating FRB source.
\section{Refraction}
\label{Refract}
Plasma refraction \citep{EM22,Z22} is strongly chromatic, like the observed
phase delays of FRB 180916B \citep{B22}.  Plasma lensing may introduce
frequency-dependent delays.  This is analogous to plasma dispersion, but is
not proportional to the dispersion measure.  A thin lens far from the source
may introduce delays as a result of differences between the lengths of ray
paths, straight-line in the limit of high frequency and bent at lower,
refracted, frequencies.  These delays occur even if the rays propagate in
vacuum outside thin lenses.  They may be much greater than the dispersion
delay of the lenses themselves.

The delay depends on the details of the lens geometry, but the
characteristic angle of refraction
\begin{equation}
	\theta = {\cal O}\left({\omega_p^2 \over \omega^2}\right),
\end{equation}
where $\omega_p$ is the plasma frequency and $\omega$ the frequency of the
wave.  Then the delay \citep{K22a}
\begin{equation}
	\label{refract}
	\Delta t = {\cal O} \left({D \over 2c}
	{\omega_p^2 \over \omega^2}\right),
\end{equation}
where $D$ is the distance of the lens from the source (the possibility of
such a lens close to the observer is excluded by the absence of such effects
in other FRB or in Galactic pulsars).  This form resembles that of the
ordinary plasma dispersion delay but with the lens thickness $d$ replaced by
$D$.  If $D \gg d$ the bent path delay far exceeds the dispersive delay; to
explain FRB 180916B this would have to be by a factor $\sim 10^5$.

The frequency scaling of Eq.~\ref{refract} is contradicted in FRB 180916B by
the observation \citep{B22} that the exponent of $\omega$ is $-0.23 \pm
0.05$ rather than $-2$, and by the fact that plausible values of $\omega_p$
would imply implausibly large (outside any possible host galaxy) values of
$D$.  Alternative explanations must be sought for the frequency-dependent
chromaticity of FRB 180916B.
\section{Millisecond Time Scales}
Many, if not all, bursts from repeating FRB show a drift to lower frequency
distinct from the effect of plasma dispersion, the ``sad trombone'' effect
\citep{M18,G18,C19,H19,J19,ZP21,B22}.  None have shown a ``happy trombone''
(drift to higher frequency).  This intrinsic property of bursts from
repeating FRB excludes the possibility that their bursts are produced by the
sweep of a narrow steady beam across the direction to the observer because
that would be equally likely to produce upward drifts in frequency as
downward drifts.

That hypothesis would also require that the opening angle of the beam
$\theta \lesssim 2\pi \delta t/P$, where $\delta t$ is the burst width and
$P$ the period of its rotation on the sky, presumably the observed period,
which is 16.3 d for FRB 180916B.  Then the Lorentz factor of the radiating
particles would have to satisfy $\gamma \gtrsim 1/\theta \sim 10^8$, which
would be difficult to reconcile with plausible accelerating mechanisms and
losses by mechanisms such as Compton scattering.  The dominance of sad
trombones and the absence or rarity of happy trombones establishes that this
behavior and the temporal structure and widths of bursts must be
consequences of their radiation mechanism. 
\section{Day Time Scales}
Periods many orders of magnitude longer than burst durations may be
attributed to motion of an envelope within which bursts may be directed.
As argued in Sec.~\ref{symmetry}, in many, perhaps all, models, a particular
source may be time-asymmetric and chromatic, and persistently so in the same
sense of asymmetry, while the entire population of periodically modulated
repeating FRB may, or may not, all have the same sense of time-asymmetry.

Here I suggest discriminating among models on the basis of whether they
imply that all periodically modulated FRB have the {\it same\/} sense of
time-asymmetric chromaticity.  Discrimination would not be based on that
sense, that is difficult to predict from models (other than the refractive
model discarded in Sec.~\ref{Refract}), but on the universality, or lack of
universality, or a single sense.
\subsection{Statistically Symmetric Models}
In an intrinsically symmetric model a periodically modulated repeating FRB
may have either sense of time asymmetry, and these occur with equal
frequency in the FRB population.  If we run a finger along the blades of
saws oriented isotropically, we will feel gradually rising sawtooths
followed by abrupt drops as often as the opposite.  Astronomical examples
include pulsar pulses if their underlying radiation pattern is stationary on
the time scales of the pulses, and there is no preferred sense of rotation
with respect to the angular radiation pattern.  The beam pattern may be
intrinsically skew, like a sawtooth, but the rotation direction is equally
likely to produce positive time-skewness as negative time-skewness.

Statistically symmetric models of periodically modulated FRB activity include
precession (free or driven by external torques) of a rotating neutron star
\citep{LBB20,LZ21,WZW22} and slow rotation of a neutron star
\citep{BWM20,LZ21,X21}.  These are statistically symmetric if, as expected,
there is no preferential orientation of the precession or rotation rate
vector with respect to the beam pattern.
\subsection{Statistically Asymmetric Models}
In a statistically asymmetric model one sense of time-asymmetric
chromaticity is intrinsically preferred.  Examples of statistically
asymmetric phenomena include the motion of ratchets, the sounds of machine
guns, earthquakes followed by aftershocks
(but only rarely preceded by foreshocks, in contrast to the model
\citep{K86} now known as Self-Organized Criticality \citep{A11} in which the
opposite is true) and, in astronomy, the repetitive outbursts of recurrent
nov\ae\ or Solar flares that rise abruptly but decay more gradually,
reddening as they decay.

Statistically asymmetric models include those in which a compact object
interacts with an accretion disc or a stellar wind because the beginning of
such interaction (entry of the compact object or its beam into denser
matter) is intrinsically different from its end (emergence from the denser
matter), and orbital motion makes a compact object's wake in a stellar wind
lag behind, rather than advancing in front of, the compact object.

Such models have been developed by \citet{IZ20,LYW21,WIZ21}.  Precessing
accretion disc/jet models \citep{K20,K21,K22b} may be intrinsically
asymmetric because Newtonian mechanics makes the direction of precession
opposite to that of the disc's rotation, so that on one side of the
accretion funnnel the (slow) precession and (fast) rotational speeds add
while on the other side they subtract.  This may be confused by jitter
around the mean precession, and it is unclear that the asymmetry of an
accretional funnel (and jet, if there is one) is large enough to be
significant.  Lens-Thirring precession \citep{S21} is intrinsically
asymmetric in the opposite sense because it is parallel to the disc's
angular momentum; in a binary this is expected to be aligned with the
orbital angular momentum.
\section{Discussion}
The large values of $D$ implied by Eq.~\ref{refract} imply that chromaticity
on time scales of days must be intrinsic to the source rather than resulting
from propagation delays.  In the one known example, FRB 180916B, the
mechanism that modulates its activity must break the symmetry between
earlier and later phases of activity.

When more periodically modulated FRB sources are observed it will be
possible to distinguish statistically symmetric models in which the sign of
the exponent of the variation of phase with frequency is equally likely to
be either positive or negative from statistically asymmetric models in which
it must be negative, as observed in FRB 180916B (models in which it can only
be positive are excluded by that one source).  Several suggested models
appear to have been excluded even by the extant data \citep{WZW22}.  
\section*{Data Availability Statement}
This theoretical study did not obtain any new data.
\section*{Acknowledgement}
It is an honor to acknowledge a long-ago discussion with the late
oceanographer Walter Munk in which he called my attention to \citet{HMM63}.

\label{lastpage} 
\end{document}